\documentclass{aa}
\usepackage{graphicx}
\usepackage[varg]{txfonts}
\usepackage{natbib}
\bibpunct{(}{)}{;}{a}{}{,} 
\begin{document}
   \title{Fundamental properties of nearby single early B-type 
   stars\thanks{Based on observations collected at the
   Centro Astron\'omico His\-pano Alem\'an (CAHA) at Calar Alto, operated 
   jointly by the Max-
   Planck Institut f\"ur Astronomie and the Instituto de Astrof\'isica
   de Andaluc\'ia (CSIC), proposals H2001-2.2-011
   and H2005-2.2-016.
   Based on observations collected at the European Southern
   Obser\-vatory, Chile, ESO 074.B-0455(A).
   Based on spectral data retrieved from the ELODIE archive at
   Ob\-servatoire de Haute-Provence (OHP).
   Based on observations made with the Nordic Optical
   Telescope, operated
   on the island of La Palma jointly by Denmark, Finland, Iceland,
   Norway, and Sweden, in the Spanish Observatorio del Roque de los
   Muchachos of the Instituto de Astrof\'isica de Canarias.}}

   \author{Mar\'ia-Fernanda Nieva\inst{1,2}\fnmsep\thanks{Visiting scientist at OAC
	  \& IATE, UNC, C\'ordoba, Argentina.}
	  \and
	  Norbert Przybilla\inst{2}\fnmsep\thanks{Visiting professor at OAC \& IATE,
	  UNC, C\'ordoba, Argentina.}
	  }

   \institute{
   $^1$ Dr.~Karl Remeis-Observatory\,\&\,ECAP,
   University Erlangen-Nuremberg, Sternwartstr.\,7, D-96049 Bamberg, Germany\\
   $^2$ Institute for Astro- and Particle Physics, University of
   Innsbruck, Technikerstr.~25/8, A-6020 Innsbruck, Austria\\
   \email{Maria-Fernanda.Nieva@sternwarte.uni-erlangen.de;norbert.przybilla@uibk.ac.at}
            }

   \date{Received; accepted}

 
  \abstract
   {}
   {Fundamental parameters 
   of a sample of 26 apparently slowly-rotating single early B-type stars 
   in the solar neighbourhood
   are presented and compared to 
   high-precision data from detached eclipsing binaries (DEBs). 
   Together with surface abundances for light elements the data 
   are used to discuss the evolutionary status of the stars in context
   of the most recent Geneva grid of models for core
   hydrogen-burning stars in the mass-range $\sim$6 to 18\,$M_\odot$ at
   metallicity $Z$\,=\,0.014.}
   {The fundamental parameters are derived on the basis of accurate and precise
   atmospheric parameters determined earlier by us from
   non-LTE analyses of high-quality spectra of the sample stars, 
   utilising the new Geneva stellar evolution models.}
   {Evolutionary masses plus radii and luminosities are determined to better 
   than typically 5\%, 10\%, and 20\% uncertainty, respectively,
   facilitating the mass-radius and mass-luminosity relationships to be 
   recovered for single core hydrogen-burning objects with a similar 
   precision as derived from DEBs. Good agreement between evolutionary and
   spectroscopic masses is found. Absolute visual and bolometric
   magnitudes are derived to typically $\sim$0.15-0.20\,mag uncertainty.
   Metallicities are constrained to
   better than 15-20\% uncertainty and tight constraints on evolutionary
   ages of the stars are provided. 
   Overall, the spectroscopic distances and ages of individual 
   sample stars agree with independently derived values for the host
   OB associations. Signatures of mixing 
   with CN-cycled material are found in 1/3 of the sample stars.
   Typically, these are
   consistent with the amount predicted by the new
   Geneva models with rotation. The presence of magnetic fields 
   appears to augment the mixing efficiency. In addition, a few
   objects are possibly the product of binary evolution. In
   particular, the unusual
   characteristics of $\tau$\,Sco 
   point to a blue straggler nature, due to a binary merger.}
   {The accuracy and precision achieved in the determination of
   fundamental stellar parameters from the quantitative
   spectroscopy of single early B-type stars comes close 
   (within a factor 2--4) to data derived from DEBs.
   While our fundamental parameters are in good agreement with those
   derived from DEBs as a function of spectral type, significant systematic 
   differences with data from the astrophysical reference literature are found.
   Masses are $\sim$10-20\% and radii $\sim$25\% lower then the recommended
   values for luminosity class V, resulting in the stars being 
   systematically fainter than assumed usually, by $\sim$0.5\,mag in 
   absolute visual and bolometric magnitude. Our sample of giants is too small
   to derive firm conclusions, but similar trends as for the dwarfs
   are indicated.}

	       
   \authorrunning{Nieva \& Przybilla}
   \titlerunning{Fundamental properties of nearby single early B-type stars}

   \maketitle
%

\section{Introduction}
Massive stars are main drivers of the evolution of galaxies
because of their energy and momentum input into the interstellar
medium (ISM)
through stellar winds and supernovae, and they are important
sites of nucleosynthesis. The development of a detailed understanding
of their inner structure and evolution is of utmost importance for wide 
fields of astrophysics. Among many others, evolution models of massive
stars provide the basis for the interpretation of stellar
populations in star clusters and galaxies \citep[e.g.][]{MaCo94}. 
They describe the structure of supernova progenitors and allow chemical yields 
to be predicted \citep[][]{Hirschietal04,ChLi13}. 
They are the starting point for explaining exotic phenomena such as neutron 
stars, stellar black holes and $\gamma$-ray bursts of the long-duration
soft-spectrum type \citep[][]{WoHe12}. 

Enormous progress has been made in the modelling of massive star 
evolution by accounting for the effects of rotation 
\citep[][]{HeLa00,MeMa00,MeMa05} and 
-- to date only far less comprehensively investigated -- 
the effects of interplay of rotation and magnetic fields 
\citep[][]{Hegeretal05,MaMe05,Meynetetal11}. 
There is a growing body of evidence that rotation together with mass loss are
physical key ingredients shaping the evolution of massive stars
throughout cosmic history \citep[for a recent review see][]{MaMe12}, 
as well as mass transfer in close binary systems
\citep{Vanbeverenetal98,Wellsteinetal01,Langer12}.

However, it is not clear whether the physical processes currently accounted for 
are comprehensive and sufficiently well represented in the stellar 
evolution models. The current models make detailed 
predictions on the surface properties of massive stars (atmospheric parameters 
and elemental abundances) as a function of mass, initial chemical composition and 
initial rotation rate. This facilitates the assumptions made in the
models to be verified, by comparison to detailed  observations.

Comprehensive tests of stellar evolution models require as accurate
characterisation of all stellar properties as possible. Besides
atmospheric parameters like effective temperature $T_\mathrm{eff}$ and
(logarithmic) surface gravity $\log g$ and elemental abundances 
a knowledge of fundamental stellar parameters mass $M$, radius $R$ and
luminosity $L$, and of age $\tau$ is necessary. Primary source of such data are 
double-lined detached eclipsing binaries (DEBs), which allow a direct 
determination of accurate masses and radii at a precision of 
1--2\% from analysis of the Keplerian orbits. As only the
$T_\mathrm{eff}$-ratio of the two stars in a DEB is tightly constrained 
from light-curve and radial-velocity-curve analysis, but not the effective temperatures 
of the components, stellar 
luminosities remain slightly less well-constrained. Ages need to be
derived by comparison with theoretical isochrones, with particular
constraints set by the condition that both components have to be coeval.
The most accurate and precise fundamental parameters for massive stars
available at present can be found in the compilation by
\citet[henceforth abbreviated TAG10]{Torresetal10}, mainly for the most common among them,
the early B-type stars of spectral types B0 to B3.
Data on abundances of individual chemical elements in massive DEBs are scarce,
though ongoing projects promise to alleviate the situation 
\citep{PaHe05,PaSo09,Pavlovskietal09,Pavlovskietal11,Mayeretal13,Tkachenkoetal14,PaSo14}.

Yet, the number of DEBs is limited, and it would be highly valuable if
data of high quality could be obtained for single stars.
The best candidate stars are located in clusters or associations,
where distances (and therefore luminosities) and ages can be constrained from 
photometry and main-sequence fitting. 
Despite the importance of this basic data in the astrophysical context,
little work on masses, radii, luminosities and ages of early B-type stars 
has been published over the past two decades 
\citep[but all far less accurate and precise than the DEB
data]{Wolff90,GiLa92,Kilian92,Lyubimkovetal02,Hohleetal10}. 
Moreover, the astrophysical reference literature like the Landolt-B\"ornstein
\citep{Schmidt-Kaler82}, which has also been adopted in more recent
compilations like Allen's Astrophysical Quantities \citep{Cox00}, are 
based on the state-of-the art achieved more than three decades ago.

Over the past years, we have improved the modelling and analysis of
the atmospheres of early B-stars by introducing non-LTE
line-formation calculations based on a new generation of sophisticated
model atoms \citep{NP06,NP07,NP08,Przybillaetal08}. Quantitative
analyses of two samples of stars based on these models
\citep[henceforth abbreviated NS11 and NP12, respectively]{NS11,NP12},
provide a highly accurate and precise characterisation of their
atmospheric parameters and chemical abundances. 

Here, we want to discuss in particular the fundamental parameters $M$, $R$, $L$
and $\tau$ of the
sample stars. As these are (effectively) single stars, stellar evolution models need to be 
employed in their derivation. Among the available state-of-the-art
stellar model grids we focus on the new generation of Geneva models  
\citep[EGE12]{Ekstroemetal12}, as these have been computed for the
same metallicity as found for the early B-stars in the solar
neighbourhood (NP12), $Z$\,=\,0.014\footnote{The differences between
the abundance mix adopted by EGE12, i.e.~the revised solar
values by \citet{AGSS09}, and the one found for the
nearby early B-stars by NP12 are irrelevant in this context.}. 
Similar grids by other groups
differ either significantly in metalli\-city \citep[$Z$\,=\,0.0088]{Brottetal11}
or do not provide information on surface abundances of the light
elements \citep{ChLi13}, which are required for the discussion of the
evolutionary status of the stars. 
A detailed comparison of evolution tracks for single massive stars 
computed with six different state-of-the-art codes has been undertaken
recently by \citet{MaPa13}. Their conclusions support our approach, as they
find that the different models agree well for the main-sequence
evolution. The only decisive factor for the loci of the evolution
tracks in the Hertzsprung-Russell diagram (HRD) is the metallicity,
while all other factors like e.g.~the overshooting parameter have
little effect, except close to the terminal-age main-sequence.
Once the fundamental parameters are derived, they can --
together with atmospheric parameters and light element abundances -- 
also be employed to discuss the evolutionary status of the individual
objects and their accordance with the predictions from the models.

The paper is structured in the following way. A brief summary of the
quantitative spectroscopy of the sample stars is given in the next section.
Details and results of the fundamental parameter determination are
discussed in Sect.~\ref{sectfundamental}. Then, the evolutionary status 
of the sample stars is discussed in Sect.~\ref{sectevolstat}.
Functional relationships of the fundamental parameters in dependence on spectral 
type are compared to data from the reference literature in
Sect.~\ref{sectfunctional}, which is followed by some concluding remarks.


\section{Quantitative spectroscopy of the sample stars}
The present study relies on the quantitative model atmosphere analysis 
of 26 early B-type stars published earlier 
by us (NS11; NP12). All sample stars are located in nearby OB associations 
(Cas-Tau, Sco-Cen, Lac\,OB1, Ori\,OB1) or in the field
at distances of less than about 400\,pc from the Sun, see Figs.~1 and 13 of NP12 for the
spatial distribution of the stars. They are 
apparently slow rotators, with projected rotational velocities 
$v_\mathrm{rot}\sin i$\,$\lesssim$\,30\,km\,s$^{-1}$.

High-resolution \'echelle spectra at very high signal-to-noise 
ratio ($S/N$\,$\simeq$\,250--800 in $B$) and wide wavelength 
coverage were investigated. The observational data for stars 1--6 (see
Table~\ref{bigtable}) were obtained with {\sc Feros} 
on the ESO 2.2m telescope in La Silla/Chile (resolving power
$R$\,$=$\,$\lambda/\Delta\lambda$\,$\approx$\,48\,000), for stars
7--16 with {\sc Foces}
on the 2.2m telescope at Calar Alto/Spain ($R$\,$\approx$\,40\,000),
and for stars 21--26 with {\sc Fies} on the 2.5m Nordic Optical
Telescope/La Palma ($R$\,$\approx$\,46\,000). The spectra for 
stars 17--20 were extracted from the {\sc Elodie} 
archive \citep[$R$\,$\approx$\,42\,000]{Moultakaetal04}. Details on the
observations and data reduction are provided by NS11 and NP12.

The quantitative analysis was carried out following a hybrid non-LTE approach
(i.e.~deviations from local thermodynamic equilibrium, LTE, were
accounted for) discussed by \citet{NP07,NP08,NP12} and \citet{PNB11}. 
In brief, line-blanketed LTE model atmospheres were computed with
{\sc Atlas9} \citep{Kurucz93} and non-LTE line-formation             
calculations were performed using updated and extended 
versions of {\sc Detail} and {\sc Surface} \citep{Giddings81,BuGi85}.
In the following we abbreviate our approach by `ADS', according to the
initials of the three involved codes.
State-of-the-art model atoms were adopted (see Table~3 of NP12 for
details) which allowed atmospheric parameters and elemental abundances 
to be obtained with high accuracy and precision. 

\begin{table*}[ht!]
\centering
\caption[]{Properties of the program stars.\\[-6mm]\label{bigtable}}
\footnotesize
\setlength{\tabcolsep}{.09cm}
 \begin{tabular}{rrl@{\hspace{-.5mm}}r@{\hspace{.2mm}}r@{\hspace{6mm}}rrcc@{\hspace{2mm}}rlrrrrrcr}
 \noalign{}
\hline
\hline
\#&Star & Sp.\,Type\tablefootmark{a} & & $V$\tablefootmark{a} & $T_\mathrm{eff}$ &$\log\,g$ &  
$\varepsilon$($\sum$CNO)\tablefootmark{b} & $Z$\tablefootmark{c} & 
$M_\mathrm{evol}$ & $R$ & $\log L/L_\odot$ & 
$M_\mathrm{spec}$ & $\log \tau_\mathrm{evol}$ & $M_V^0$ & $M_\mathrm{bol}$ &
mixing\tablefootmark{d} & $B$\\ 
& &   & & mag & K    & (cgs)   &
& & $M_\odot$ & $R_\odot$ & & $M_\odot$ & (yr) & mag & mag & status & G\\[-.2mm]
\hline\\[-3mm]
1 & \object{HD36591} & B1\,V     &     & 5.339 & 27000 & 4.12 &  8.92 & 0.0138 & 11.8 & 5.1 & 4.10 & 12.6 & 6.87   & $-$2.85 & $-$5.49 & u &                 {\ldots}\\
  &                  &           &$\pm$& \ldots&   300 & 0.05 &  0.07 & 0.0025 &  0.3 & 0.4 & 0.06 &  2.8 & $^{+0.06}_{-0.09}$   &    0.14 &    0.14 &   &           \\
2 & \object{HD61068} & B1\,V     &     & 5.711 & 26300 & 4.15 &  8.94 & 0.0141 & 11.1 & 4.8 & 3.99 &\ldots& 6.87   & $-$2.65 & $-$5.24 & m &                 {\ldots}\\
  & ~PT\,Pup          &           &$\pm$& 0.015 &   300 & 0.05 &  0.06 & 0.0026 &  0.3 & 0.4 & 0.06 &\ldots& $^{+0.08}_{-0.15}$   &    0.15 &    0.15 &   &           \\
3 & \object{HD63922} & B0.2\,III &     & 4.106 & 31200 & 3.95 &  8.95 & 0.0144 & 18.1 & 7.7 & 4.70 &\ldots& 6.78   & $-$4.08 & $-$7.01 & u &                 {\ldots}\\
  & ~P\,Pup          &           &$\pm$& 0.005 &   300 & 0.05 &  0.07 & 0.0024 &  0.6 & 0.5 & 0.06 &\ldots& $^{+0.01}_{-0.03}$   &    0.14 &    0.14 &   &           \\
4 & \object{HD74575} & B1.5\,III &     & 3.679 & 22900 & 3.60 &  8.97 & 0.0146 & 12.1 & 9.4 & 4.34 & 10.0 & 7.17   & $-$3.89 & $-$6.11 & u &                      -- \\
  & ~$\alpha$\,Pyx   &           &$\pm$& 0.006 &   300 & 0.05 &  0.06 & 0.0028 &  0.6 & 0.7 & 0.06 &  2.4 & $^{+0.03}_{-0.02}$   &    0.15 &    0.15 &   &       (1) \\
5 &\object{HD122980} & B2\,V     &     & 4.353 & 20800 & 4.22 &  8.90 & 0.0127 &  7.1 & 3.5 & 3.32 &  8.7 & 7.08   & $-$1.50 & $-$3.56 & u &                 {\ldots}\\
  & ~$\chi$\,Cen     &           &$\pm$& 0.007 &   300 & 0.05 &  0.04 & 0.0020 &  0.2 & 0.3 & 0.06 &  2.0 & $^{+0.13}_{-0.23}$   &    0.14 &    0.15 &   &           \\
6 &\object{HD149438} & B0.2\,V   &     & 2.825 & 32000 & 4.30 &  8.97 & 0.0152 & 15.5 & 4.7 & 4.33 & 17.4 & $<$6.3:& $-$3.02 & $-$6.08 & m &               $\sim$500 \\
  & ~$\tau$\,Sco     &           &$\pm$& 0.009 &   300 & 0.05 &  0.06 & 0.0027 &  0.6 & 0.3 & 0.06 &  6.3 & \ldots &    0.14 &    0.15 &   &                     (2) \\
7 & \object{HD886}   & B2\,IV    &     & 2.834 & 22000 & 3.95 &  8.92 & 0.0137 &  8.8 & 5.4 & 3.78 &  9.8 & 7.32   & $-$2.51 & $-$4.71 & u &                      -- \\
  & ~$\gamma$\,Peg   &           &$\pm$& 0.015 &   400 & 0.05 &  0.07 & 0.0021 &  0.3 & 0.4 & 0.06 &  3.6 & $^{+0.02}_{-0.03}$   &    0.15 &    0.16 &   &     (3,4) \\
8 & \object{HD29248} & B1.5\,IV  &     & 3.930 & 22000 & 3.85 &  8.95 & 0.0143 &  9.3 & 6.2 & 3.90 &  8.6 & 7.32   & $-$2.84 & $-$5.02 & m &                      -- \\
  & ~$\nu$\,Eri      &           &$\pm$& 0.023 &   250 & 0.05 &  0.06 & 0.0026 &  0.3 & 0.5 & 0.06 &  2.0 &  $^{+0.02}_{-0.02}$  &    0.15 &    0.15 &   &       (3) \\
9 & \object{HD35299} & B1.5\,V   &     & 5.694 & 23500 & 4.20 &  8.99 & 0.0154 &  8.7 & 4.0 & 3.64 &  6.0 & 6.92   & $-$1.98 & $-$4.36 & u &                 {\ldots}\\
  &                  &           &$\pm$& 0.010 &   300 & 0.05 &  0.06 & 0.0026 &  0.3 & 0.3 & 0.06 &  2.6 &  $^{+0.13}_{-0.22}$  &    0.14 &    0.15 &   &           \\
10& \object{HD35708} & B2\,V     &     & 4.875 & 20700 & 4.15 &  9.01 & 0.0161 &  7.3 & 3.9 & 3.39 &  8.1 & 7.25   & $-$1.69 & $-$3.74 & m &                 {\ldots}\\
  & ~$o$\,Tau        &           &$\pm$& 0.012 &   200 & 0.07 &  0.07 & 0.0025 &  0.3 & 0.4 & 0.08 &  2.2 & $^{+0.08}_{-0.15}$   &    0.20 &    0.20 &   &           \\
11& \object{HD36512} & B0\,V     &     & 4.618 & 33400 & 4.30 &  8.93 & 0.0140 & 17.2 & 5.0 & 4.45 & 22.8 & $<$6.3 & $-$3.22 & $-$6.38 & u &                 {\ldots}\\
  & ~$\upsilon$\,Ori &           &$\pm$& 0.013 &   200 & 0.05 &  0.07 & 0.0025 &  0.6 & 0.4 & 0.06 &  8.1 & \ldots &    0.14 &    0.14 &   &                         \\
12& \object{HD36822} & B0.5\,III &     & 4.408 & 30000 & 4.05 &  8.88 & 0.0129 & 15.5 & 6.3 & 4.46 & 16.7 & 6.78   & $-$3.53 & $-$6.42 & m &                 {\ldots}\\
  & ~$\phi^1$\,Ori   &           &$\pm$& 0.006 &   300 & 0.10 &  0.07 & 0.0023 &  1.1 & 1.0 & 0.12 &  7.8 & $^{+0.06}_{-0.15}$   &    0.30 &    0.30 &   &           \\
13& \object{HD36960} & B0.7\,V   &     & 4.785 & 29000 & 4.10 &  8.87 & 0.0130 & 13.7 & 5.6 & 4.30 & 15.9 & 6.78   & $-$3.18 & $-$6.01 & u &                      -- \\
  &                  &           &$\pm$& 0.007 &   300 & 0.07 &  0.05 & 0.0023 &  0.7 & 0.6 & 0.08 &  6.1 & $^{+0.07}_{-0.14}$  &    0.20 &    0.21 &   &        (5) \\
14&\object{HD205021} & B1\,IV    &     & 3.233 & 27000 & 4.05 &  8.87 & 0.0132 & 12.2 & 5.6 & 4.18 & 19.5 & 6.94   & $-$3.03 & $-$5.70 & m &$\sim$300\tablefootmark{e}\\
  & ~$\beta$\,Cep    &           &$\pm$& 0.014 &   450 & 0.05 &  0.07 & 0.0025 &  0.4 & 0.4 & 0.06 &  6.5 & $^{+0.04}_{-0.05}$  &    0.15 &    0.15 &   &        (6) \\
15&\object{HD209008} & B3\,III   &     & 5.995 & 15800 & 3.75 &  8.96 & 0.0142 &  5.8 & 5.5 & 3.22 &\ldots& 7.80   & $-$1.98 & $-$3.32 & u &                 {\ldots}\\
  & ~18\,Peg         &           &$\pm$& 0.008 &   200 & 0.05 &  0.07 & 0.0026 &  0.4 & 0.5 & 0.07 &\ldots& $^{+0.10}_{-0.03}$  &    0.17 &    0.17 &   &            \\
16&\object{HD216916} & B1.5\,IV  &     & 5.587 & 23000 & 3.95 &  8.94 & 0.0142 &  9.5 & 5.6 & 3.89 &  8.6 & 7.25   & $-$2.67 & $-$4.99 & u &                 {\ldots}\\
&   ~EN\,Lac         &           &$\pm$& 0.015 &   200 & 0.05 &  0.05 & 0.0024 &  0.3 & 0.4 & 0.06 &  2.2 & $^{+0.02}_{-0.02}$   &    0.14 &    0.15 &   &           \\
17&  \object{HD3360} & B2\,IV    &     & 3.661 & 20750 & 3.80 &  9.00 & 0.0159 &  8.7 & 6.3 & 3.82 &  8.6 & 7.40   & $-$2.77 & $-$4.82 & m &$\sim$335\tablefootmark{e}\\
  & ~$\zeta$\,Cas    &           &$\pm$& 0.017 &   200 & 0.05 &  0.05 & 0.0024 &  0.3 & 0.5 & 0.06 &  1.7 & $^{+0.01}_{-0.03}$   &    0.14 &    0.15 &   &       (7) \\
18& \object{HD16582} & B2\,IV    &     & 4.067 & 21250 & 3.80 &  8.98 & 0.0150 &  9.1 & 6.4 & 3.87 &  6.5 & 7.36   & $-$2.82 & $-$4.93 & m &                      -- \\
  & ~$\delta$\,Cet   &           &$\pm$& 0.007 &   400 & 0.05 &  0.05 & 0.0024 &  0.3 & 0.5 & 0.06 &  1.6 & $^{+0.03}_{-0.03}$   &    0.15 &    0.15 &   &     (1,3) \\
19& \object{HD34816} & B0.5\,V   &     & 4.286 & 30400 & 4.30 &  8.91 & 0.0141 & 13.8 & 4.5 & 4.19 & 15.0 & $<$6.3 & $-$2.77 & $-$5.73 & u &                 {\ldots}\\
  & ~$\lambda$\,Lep  &           &$\pm$& 0.005 &   300 & 0.05 &  0.06 & 0.0024 &  0.4 & 0.3 & 0.06 &  3.5 & \ldots &    0.15 &    0.15 &   &                         \\
20&\object{HD160762} & B3\,IV    &     & 3.800 & 17500 & 3.80 &  8.98 & 0.0149 &  6.6 & 5.5 & 3.41 &  5.6 & 7.65   & $-$2.16 & $-$3.78 & u &                      -- \\
  & ~$\iota$\,Her    &           &$\pm$& 0.000 &   200 & 0.05 &  0.06 & 0.0025 &  0.2 & 0.4 & 0.06 &  1.0 & $^{+0.02}_{-0.02}$   &    0.14 &    0.14 &   &       (8) \\
21& \object{HD37020} & B0.5\,V   &     & 6.720 & 30700 & 4.30 &  8.99 & 0.0153 & 14.0 & 4.5 & 4.21 & 15.3 & $<$6.4 & $-$2.80 & $-$5.79 & u &                 {\ldots}\\
  & ~$\theta^1$\,Ori\,A&         &$\pm$& \ldots&   300 & 0.08 &  0.05 & 0.0024 &  0.8 & 0.5 & 0.09 &  3.5 & \ldots &    0.23 &    0.23 &   &                         \\
22& \object{HD37042} & B0.5\,V   &     & 6.380 & 29300 & 4.30 &  8.94 & 0.0146 & 12.8 & 4.3 & 4.09 & 14.8 & $<$6.3 & $-$2.60 & $-$5.49 & m &                      -- \\
  &  ~$\theta^2$\,Ori\,B&        &$\pm$& \ldots&   300 & 0.05 &  0.06 & 0.0025 &  0.6 & 0.4 & 0.07 &  3.4 & \ldots &    0.18 &    0.19 &   &                     (8) \\
23& \object{HD36959} & B1.5\,V   &     & 5.670 & 26100 & 4.25 &  8.94 & 0.0147 & 10.3 & 4.1 & 3.85 & 13.8 & 6.48   & $-$2.24 & $-$4.87 & u &                 {\ldots}\\
  &                  &           &$\pm$& \ldots&   200 & 0.07 &  0.08 & 0.0027 &  0.5 & 0.4 & 0.08 &  5.2 & $^{+0.31}_{...}$   &    0.20 &    0.20 &   &             \\
24& \object{HD37744} & B1.5\,V   &     & 6.213 & 24000 & 4.10 &  8.91 & 0.0137 &  9.5 & 4.7 & 3.81 &  7.6 & 7.10   & $-$2.37 & $-$4.80 & u &                 {\ldots}\\
  &                  &           &$\pm$& 0.011 &   400 & 0.10 &  0.06 & 0.0021 &  0.4 & 0.7 & 0.12 &  2.6 & $^{+0.08}_{-0.20}$   &    0.29 &    0.29 &   &           \\
25& \object{HD36285} & B2\,V     &     & 6.315 & 21700 & 4.25 &  8.97 & 0.0148 &  7.5 & 3.5 & 3.39 &  9.6 & 6.85   & $-$1.54 & $-$3.73 & u &                 {\ldots}\\
  &                  &           &$\pm$& 0.008 &   300 & 0.08 &  0.07 & 0.0025 &  0.4 & 0.4 & 0.09 &  3.9 & $^{+0.28}_{...}$   &    0.23 &    0.23 &   &             \\
26& \object{HD35039} & B2\,IV    &     & 4.731 & 19600 & 3.56 &  8.94 & 0.0138 &  9.0 & 8.5 & 3.98 &  7.8 & 7.43   & $-$3.32 & $-$5.21 & u &                 {\ldots}\\
  & ~$o$\,Ori        &           &$\pm$& 0.011 &   200 & 0.07 &  0.06 & 0.0025 &  0.5 & 0.9 & 0.08 &  2.9 & $^{+0.05}_{-0.06}$   &    0.20 &    0.21 &   &           \\
\hline
\vspace{-7mm}
\end{tabular}
\tablefoot{
\tablefoottext{a}{see \citet{Nieva13};~~}
\tablefoottext{b}{$\varepsilon(\sum\mathrm{CNO})=\log
(\Sigma\mathrm{CNO}/\mathrm{H})\,+\,12$, using ADS abundances for C, N and O
from NP12 (stars 1--20) and NS11 (stars 21--26);~~}
\tablefoottext{c}{by mass fraction;~~}
\tablefoottext{d}{m/u: atmosphere mixed/unmixed with CN-processed
material;~~}
\tablefoottext{e}{inferred polar field strength for the observed dipolar $B$-field (oblique magnetic rotator).}
}\\[-1mm]
\tablebib{(1)~\citet{Bagnuloetal12}; (2)~\citet{Donatietal06}; (3)~\citet{Silvesteretal09}; (4)~\citet{Neineretal13};
(5)~\citet{Bagnuloetal06}; (6)~\citet{Henrichsetal13}; (7)~\citet{Neineretal03}; (8)~\citet{Schnerretal08}
}
\end{table*}

Spectral types, measured $V$ magnitudes, effective temperatures $T_\mathrm{eff}$ 
and surface gravities $\log g$ (in cgs units) for the sample stars together 
with their respective uncertainties (1$\sigma$-values) are summarised in
Table~\ref{bigtable}, see Table~3 of NS11 and Tables~5 and 6 of NP12 
for further data on atmospheric parameters and chemical abundances.
Note that the derived surface gravities for the individual stars are 
practically identical to the polar gravities \citep[which is the decisive
quantity in terms of stellar evolution,][]{MaMe12}. There is 
observational evidence that many of the objects are true slow rotators, 
see Sect.~\ref{sectevolstat}. Moreover, it is likely that
among the remaining stars true fast rotators seen nearly pole-on --
only therefore showing sharp lines -- are absent.

The sample has passed a critical test for the quality of the
quantitative analysis in terms that the derived CNO abundances follow
tightly the predicted nuclear path \citep{Przybillaetal10,Maederetal14}, see Fig.~14 of NP12.
For a verification of the catalytic nature of the CNO cycles we also
provide the sum of carbon, nitrogen and oxygen abundances in Table~\ref{bigtable}. The
sample stars turn out to be homogeneous in $\varepsilon(\sum$CNO) 
on the 10\%-level, 
similar to the general trend of chemical homogeneity established in NP12.
The sample mean value is 
$\overline{\varepsilon}(\sum\mathrm{CNO})$\,=\,8.94$\pm$0.04.

Metallicities $Z$ were computed for the individual
objects on the basis of the abundance data published by NS11 and NP12.
In addition to the most abundant species discussed in these studies
(C, N, O, Ne, Mg, Si, Fe), data
for all other metals in the periodic system up to zinc were considered 
for constraining $Z$, adopting solar meteoritic values of
\citet{AGSS09}, except for chlorine and argon, where abundances from the 
Orion nebula were taken \citep{Estebanetal04}. Any deviations of these
auxiliary data from the ‘true’ cosmic values are likely to be absorbed by the
error margins of $Z$ due to their small contribution.


\section{Fundamental parameter determination \& results}{\label{sectfundamental}
\begin{figure}[!t]
\centering
\includegraphics[width=.95\linewidth]{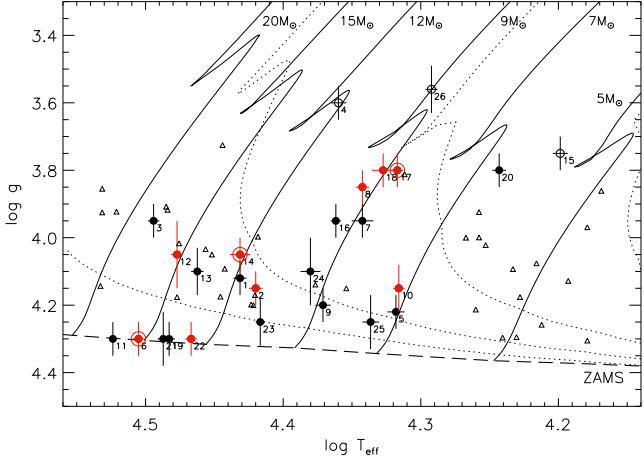}
\caption{The sample stars in the $\log T_\mathrm{eff}$--$\log g$-plane.
Black/red dots denote CN-unmixed/mixed chemical composition and open
thick circles objects near/beyond core-H exhaustion. Wide circles surrounding
the dots mark magnetic stars. Error bars denote 1$\sigma$-uncertainties.
Data from double-lined detached
eclipsing binary stars \citep{Torresetal10} are shown for comparison
(small triangles). Overplotted are evolution tracks for non-rotating
stars at $Z$\,=\,0.014 from \citet{Ekstroemetal12}, for initial masses
as indicated, and isochrones for $\log \tau_\mathrm{evol}$(yr)\,=\,6.5,
7.0, 7.5 and 8.0 (dotted lines, left to right). The position of the ZAMS 
is indicated by the long-dashed line. The sample stars are marked
according to the numbering scheme introduced in Table~\ref{bigtable}. 
}
\label{teff-logg}
\end{figure}

Once the basic atmospheric parameters $T_\mathrm{eff}$ and $\log g$
are known, evolutionary masses $M_\mathrm{evol}$ of stars can be determined by
comparison with the loci of stellar evolution tracks in the
$T_\mathrm{eff}$--$\log g$ plane. This is visualised for our sample
stars in Fig.~\ref{teff-logg}, with evolution tracks and isochrones for  
non-rotating stars adopted from EGE12. Our stars span a range of
$\sim$6 to 18\,$M_\odot$, covering the main-sequence band from the
zero-age main sequence (ZAMS) to the terminal-age main sequence
(TAMS). Among the most massive stars several objects from the youngest
parts in the Orion OB1 association and $\tau$\,Sco in the Upper Sco
association fall on the ZAMS. Three objects close to the TAMS
could either be in the last phases of core H-burning, or may have
terminated core H-burning, and have started shell H-burning already. 
The status of these stars is in
particular not clear because rotation widens the main sequence (EGE12).
Note that the $M_\mathrm{evol}$ data in Table~\ref{bigtable} deviate
slightly from the values presented in NP12, as the latter were based
on stellar evolution tracks of \citet{MeMa03} for $Z$\,=\,0.02.
In anticipation of the discussion in Sect.~\ref{sectevolstat},
additional information on the status of mixing of the stellar surface with
CN-processed material from the core and on the presence of surface
magnetic fields is encoded in Fig.~\ref{teff-logg}. 

For comparison, positions of DEB components discussed by TAG10 are
also displayed in Fig.~\ref{teff-logg}. DEB components at this point of 
evolution qualify as excellent examples of slowly-rotating single stars 
\citep{deMinketal11}. They are spherical, have not experienced mass exchange, 
and rotate at velocities of typically $\sim$100\,km\,s$^{-1}$ --
in practice they are therefore slow rotators for a comparison with
stellar evolution models. The periods of the massive
binaries discussed by TAG10 (typically 2-10\,days) may be long enough
to significantly reduce the potential effects of tides on the 
structure and evolution of the binary components \citep{Songetal13}.

\begin{figure}[!t]
\centering
\includegraphics[width=.95\linewidth]{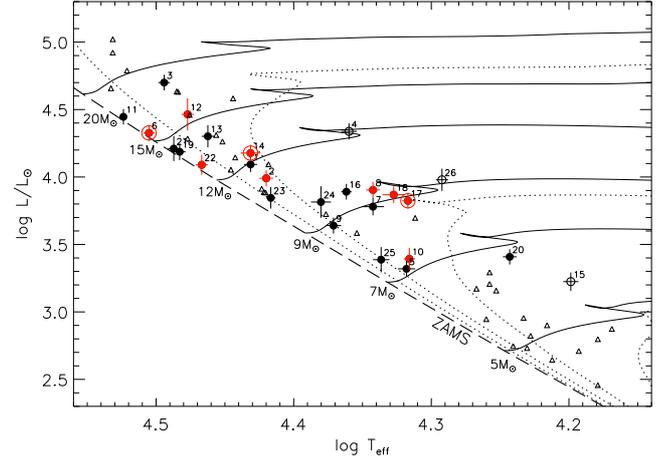}
\caption{The sample stars in the $\log T_\mathrm{eff}$-$\log L$-plane.
The symbol encoding is the same as in Fig.~\ref{teff-logg}.
}
\label{teff-logL}
\end{figure}

A representation complementary to the $\log T_\mathrm{eff}$--$\log g$
plane is the physical Hertzsprung--Russell diagram (HRD), stellar
luminosity vs.~$\log T_\mathrm{eff}$, see Fig.~\ref{teff-logL}. 
The derivation of the luminosity from measured $V$ magnitudes 
requires a knowledge of the distance and 
the extinction along the line-of-sight to each sample star. 
We follow the same strategy as in NP12 for the calculation of
spectroscopic distances, and extend it to the additional stars from NS11.
Then, bolometric corrections are adopted from \citet[N13]{Nieva13}. 
The resulting luminosities are summarised in Table~\ref{bigtable},
together with the deduced radii (from the relation between $L$,
$T_\mathrm{eff}$ and $R$).
The distribution of the sample stars relative to the stellar evolution
tracks is the same as in Fig.~\ref{teff-logg}.

Next, we put the sample stars in context of the mass-radius and the
mass-luminosity relationships as deduced from DEBs of TAG10 in Figs.~\ref{M-R}
and \ref{M-L}, respectively. We want to draw the attention to the
extremely small error bars for these particular DEB data -- both components of a DEB have
masses and radii determined to $\pm$3\%, or better. These
comprise only data of highest accuracy and precision discussed within
the much broader literature on DEBs. The predicted loci of the 
ZAMS and TAMS (non-rotating models of EGE12) are also displayed, as well as the
locus where the models reach 50\% core-H depletion. Our sample stars
fit well into the trends, with error bars coming close to that of
the DEB components, typically within a factor 2--4. The precision in
luminosity even reaches similar values as that obtained for DEBs.
This opens up the possibility to improve on the number statistics and 
in particular to trace the regions close to the ZAMS
in the $M$--$R$ and $M$--$L$ relations for the more massive objects, which so far is 
not covered by DEBs. 

\begin{figure}[!t]
\centering
\includegraphics[width=.98\linewidth]{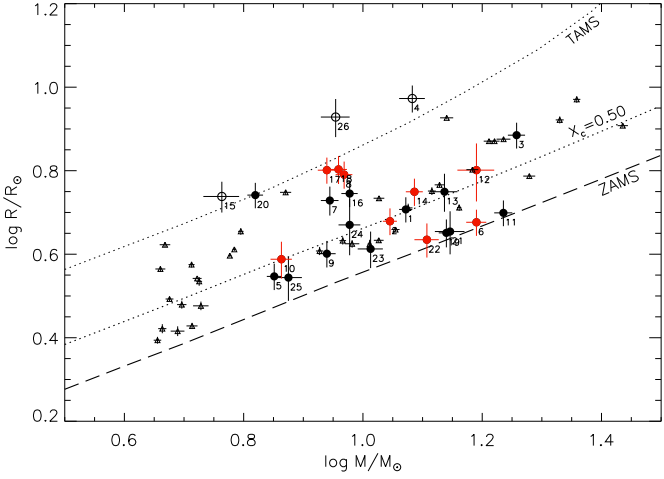}
\caption{Mass-radius relationship for the sample stars. See
Fig.~\ref{teff-logg} for the symbol encoding. Abscissa values are
evolutionary masses. In addition to the ZAMS,
two additional loci, for 50\% core-H depletion and for the TAMS, are
indicated by the thick/thin-dotted lines, as predicted by the stellar
evolution models of \citet{Ekstroemetal12}. Error bars are shown also
for the detached eclipsing binary components.
}
\label{M-R}
\end{figure}

\begin{figure}[!t]
\centering
\includegraphics[width=.98\linewidth]{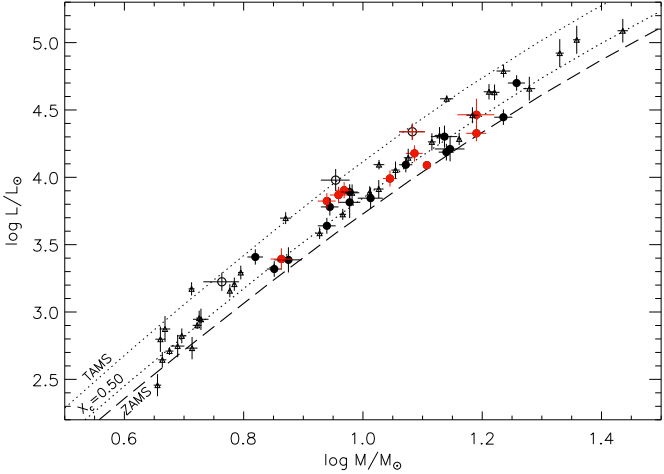}
\caption{Mass-luminosity relationship for the sample stars. See
Figs.~\ref{teff-logg} and \ref{M-R} for the symbol encoding.
}
\label{M-L}
\end{figure}

A further test for the consistency of the results is the comparison of
evolutionary with spectroscopic masses $M_\mathrm{spec}$. The latter
are derived from computing the stellar luminosity based on independent
distance determinations, either from (revised) Hipparcos parallaxes
\citep{vanLeeuwen07} or from distances to associations (see
Table~\ref{OBassoctable} for references). This allows radii to be
derived (in analogy to the procedure described above), and finally
spectroscopic masses, utilising Newton's law of gravitation.
The data on $M_\mathrm{spec}$ is summarised in
Table~\ref{bigtable}. The comparison between spectroscopic and
evolutionary masses is visualised in Fig.~\ref{M-M}. In general, there
is good agreement within the 1$\sigma$ error bars. Note that the
uncertainties in $M_\mathrm{spec}$ are large.
These are dominated by the uncertainties in the parallaxes or the association
distances, where in the latter case also systematic effects for
individual stars may occur, depending on their relative position in the radially 
extended associations.
An apparent systematic trend -- in the sense that spectroscopic 
masses appear to be larger than evolutionary masses for values
larger than about 15\,$M_\odot$ -- looses significance if two outliers are omitted.
These are HD\,36512 ($\upsilon$\,Ori, \#11) and HD\,205021
($\beta$\,Cep, \#14). In the former case, the literature value of the distance to the Ori OB1c
association may be overestimated (our spectroscopic distances for 3
other Ori OB\,1c members tend also to be systematically smaller than
the adopted association distance), while in the latter case
the object is a binary (see Sect.~\ref{sectevolstat}), which may have biased the 
parallax determination. 

\begin{figure}[!t]
\centering
\includegraphics[width=.98\linewidth]{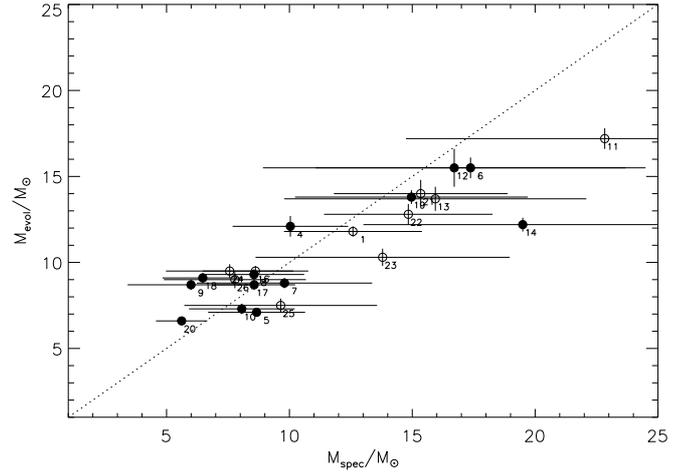}
\caption{Comparison of spectroscopic $M_\mathrm{spec}$ and
evolutionary masses $M_\mathrm{evol}$ for the sample stars.
Dots mark objects with measured parallaxes
\citep{vanLeeuwen07}, circles mark stars where
only association distances are available (see Table~\ref{OBassoctable}). 
A few field stars have no reliable parallaxes and are therefore
omitted here.
Error bars denote 1$\sigma$-uncertainties and the dotted line
indicates the 1:1 relation.
}
\label{M-M}
\end{figure}

\begin{table}[t!]
\centering
\caption[]{Comparison with OB association
properties.\\[-6mm]\label{OBassoctable}}
\footnotesize
 \begin{tabular}{llll}
 \noalign{}
\hline
\hline
Association & Star & $d$\,(pc) & $\tau$\,(Myr)\\[-.2mm]
\hline\\[-3mm]
\multicolumn{2}{l}{Upper Sco...............................} & 145$\pm$2\,(1) & $\sim$11\,(2)\\
              & \#6~~~HD149438  & 141$\pm$9& blue straggler\\[1mm]
\multicolumn{2}{l}{Upper Cen Lup.......................}& 140$\pm$2\,(1) & $\sim$16\,(2)\\
              & \#5~~~HD122980  & 146$\pm$9& 12$^{+4}_{-5}$\\[1mm]
\multicolumn{2}{l}{Lac\,OB1..................................}  & 368$\pm$17\,(1) & 12--16\,(3)\\
              & \#16~HD216916 & 398$\pm$26 & 18$\pm$1\\[1mm]
\multicolumn{2}{l}{Ori\,OB1a.................................} & $\sim$350$\pm$25\,(4) & 8--12\,(4)\\
	      & \#26~HD35039  & 388$\pm$36 & 27$^{+3}_{-4}$\\
              & \#9~~~HD35299 & 334$\pm$22 & 8$\pm$3\\[1mm]
\multicolumn{2}{l}{Ori\,OB1b.................................} & $\sim$400$\pm$15\,(4) & 5--8\,(4)\\
              & \#1~~~HD36591 & 399$\pm$25 & 7$\pm$1\\
	      & \#24~HD37744  & 462$\pm$61 & 13$^{+2}_{-5}$\\[1mm]
\multicolumn{2}{l}{Ori\,OB1c.................................} & $\sim$400$\pm$30\,(4) & 2--6\,(4)\\
              & \#25~HD36285  & 364$\pm$38 & 7$^{+6}_{{\ldots}}$\\
              & \#11~HD36512  & 358$\pm$23 & $<$2\\
	      & \#23~HD36959  & 356$\pm$32 & 3$^{+3}_{{\ldots}}$\\
	      & \#13~HD36960  & 382$\pm$36 & 6$^{+1}_{-2}$\\[1mm]
\multicolumn{2}{l}{Ori\,OB1d.................................} & 414$\pm$7\,(5) & $<$2\,(4)\\
              & \#21\,HD37020 & 408$\pm$43 & $<$3\\
	      & \#22\,HD37042 & 396$\pm$30 & $<$2\\
\hline
\vspace{-6mm}
\end{tabular}
\tablebib{(1)~\citet{deZeeuwetal99}; (2)~\citet{Pecautetal12};
(3)~\citet{Blaauw58,Blaauw91}; (4)~\citet{Bally08};
(5)~\citet{Mentenetal07}
}
\end{table}

Conservatively, one should view the uncertainties of the spectroscopic
masses, which are typically larger by a factor of five to ten than 
those for evolutionary masses,
as the limitation to the precision to which masses can be
derived for single stars. However, there is broad agreement on
evolutionary tracks for the main-sequence phase \citep[in particular
for slow- or non-rotating stars]{MaPa13}, indicating that the
evolutionary masses seem largely unaffected by systematic
errors, while distances are highly uncertain. 
We therefore argue that the small uncertainties of the evolutionary
masses may therefore be nevertheless realistic, while remaining
differences in the absolute values of evolutionary and spectroscopic
masses are due to systematic bias in the distances\footnote{We want to remind
the reader of the ongoing and unsolved discussion about
systematics affecting Hipparcos- and ground-based distance
determinations for stars as close as those in the Pleiades open cluster 
\citep[e.g.][]{vanLeeuwen09}.
Our sample objects, with a possible exception of $\gamma$\,Peg, are much more
distant.}. Without doubt, parallaxes obtained with Gaia will settle the issue.

\begin{figure}[!t]
\centering
\includegraphics[width=.95\linewidth]{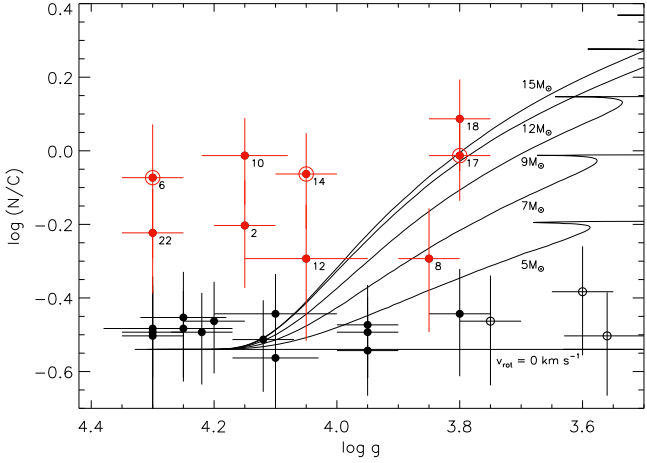}
\caption{The sample stars in the $\log g$--$\log$\,(N/C)-plane.
Symbol encoding according to Fig.~\ref{teff-logg}. Predictions from 
evolution models for rotating stars (EGE12, $Z$\,=\,0.014, for an initial
$\Omega/\Omega_\mathrm{crit}$\,=\,0.40) are indicated by full lines. 
The tracks are labeled by the corresponding ZAMS mass.
The tracks for nonrotating stars in this mass range all coincide, as
indicated.}
\label{g-NC}
\end{figure}

Finally, evolutionary ages $\tau_\mathrm{evol}$ of the sample stars
are determined by comparison with isochrones of EGE12, see
Figs.~\ref{teff-logg} and \ref{teff-logL} for a visualisation, and
Table~\ref{bigtable} for the derived data. In addition, Table~\ref{bigtable}
contains information on absolute visual $M_V^0$ and bolometric
magnitudes $M_\mathrm{bol}$ of the sample stars\footnote{Note that
in the few cases where no information is available on the uncertainty
of the $V$-magnitude (see Table~\ref{bigtable}), it has been set to zero in
the error propagation for computing the uncertainty of $M_V^0$. The
contribution of $\Delta V$ is negligible compared to the contribution
of the uncertainty in distance.}, a statement
whether the atmospheres show indications for mixing with CN-processed
material, and information on the presence and strength of a magnetic field, 
where available.

A test for the consistency of the derived properties of the sample
stars can be made for members of OB associations, which are located at
similar distances and have similar ages. Distances and ages of
OB associations are compared with the data derived here for member stars 
in Table~\ref{OBassoctable}. We have adopted uncertainties of 25\,pc,
15\,pc and 30\,pc for the distances to the Ori OB1a, OB1b and OB1c associations,
respectively, as \citet{Bally08} does not provide these data. Thus, the radial
extension of these associations is assumed to be comparable to their
extension across the sky.

Overall, good agreement is found in terms of distances and ages, i.e. within the
1$\sigma$-uncertainties. However, two distinctively discrepant objects are found. 
The first is $\tau$\,Sco in the Upper Sco association, which appears much too young
with an apparent age of $<$2\,Myr compared to the association age of
$\sim$11\,Myr \citep{Pecautetal12}. On the other hand, there is a very good mutual
agreement between spectroscopic and Hipparcos
distance (see NP12), and the association distance (see
Table~\ref{OBassoctable}). It will be shown in the next Section that
these characteristics together with other indications point towards a blue straggler
nature for this star. The second discrepant object is HD\,35039 in Ori
OB1a. The star appears too evolved (corresponding to an age discrepant
by $\sim$4$\sigma$) in comparison to other association members, despite
the distance, radial velocity and proper motion is compatible. We suggest 
that HD\,35039 belongs to the (older) field population overlapping with
the Ori OB1a association. The alternative, that it has evolved to its current 
characteristics through a binary channel \citep[it is a single-lined
spectroscopic binary, SB1, with a period of 290\,d,][ET08]{EgTo08},
appears less appealing. CN-processed material should have been transferred in the
process, which is not traced at the stellar surface.


\section{Evolutionary status of the sample stars\label{sectevolstat}}
The majority of the sample stars falls on the main sequence band (see
Figs.~\ref{teff-logg} to \ref{M-L}), i.e. they
are core hydrogen-burning stars. 
Most among them seem to have used up $\sim$50\% of their hydrogen
fuel. Whether the three objects close to the TAMS (HD35039, HD74575 and HD209008)
are still core hydrogen-burning or have already started shell
hydrogen-burning will depend on their (unknown) initial rotation
rates. This is because rotation increases the core mass,
extends the stellar main-sequence lifetime and therefore shifts the
TAMS towards lower gravities/higher luminosities (EGE12).
Figures~\ref{teff-logg} and \ref{teff-logL} therefore indicate the lower
boundary for the TAMS position, as evolution tracks for {\em
non-rotating} stars are shown.

\begin{figure}[!t]
\centering
\includegraphics[width=.95\linewidth]{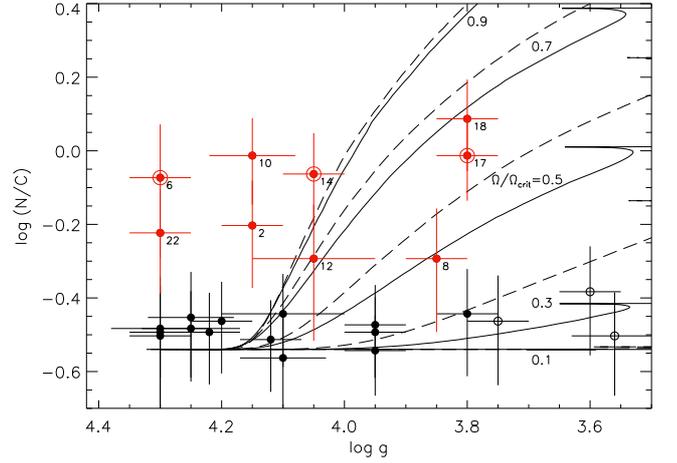}
\caption{Same as Fig.~\ref{g-NC}, but showing the effects of varying
initial rotational velocity. Full and long-dashed lines mark predictions of
\citet{Georgyetal13} 
for a 9\,$M_{\odot}$ and a 15\,$M_{\odot}$ model at $Z$\,=\,0.014,
respectively, for initial $\Omega/\Omega_\mathrm{crit}$ varying 
from 0.1 to 0.9, as indicated. 
}
\label{g-NCrot}
\end{figure}

The true rotational velocities of the sample stars are primarily not
known, as only the projected rotational velocities $v_{\rm rot} \sin i$ 
can be determined from the available optical spectra. The sample could 
therefore contain faster rotators seen at low inclination $i$. 
One possibility to distinguish fast rotators seen nearly pole-on
from slow rotators may come from the observed line-profile shapes in
high-resolution and high-S/N spectra:
fast rotators seen nearly pole-on produce flat-bottomed profiles 
with a trapezoidal appearance in some weak lines like prominently observed in Vega
\citep{Gulliveretal94} and some Be stars (T.~Rivinius,
priv.~comm.) -- such are not present in the sample star spectra.
Another possibility, motivated by evolution models for
rotating stars, may come from surface N/C ratios (see Figs.~\ref{g-NC}
and \ref{g-NCrot}). Noticeable effects on the main sequence are predicted to occur 
when OB stars evolve off the ZAMS and attain surface gravities $\log g$ in the range
$\sim$4.0 to 3.9 for ratios of initial to critical angular velocity
$\Omega$/$\Omega_\mathrm{crit}$\,$\gtrsim$\,0.4.
Therefore, even stars with an average $v_{\rm rot}$ of
$\sim$100 to 140\,km\,s$^{-1}$ (depending on mass) on the main
sequence -- corresponding to $v_{\rm rot}$ up to slightly larger than 200\,km\,s$^{-1}$ on the
ZAMS -- are expected to show a behaviour similar to that predicted by 
non-rotating models until the end of H core-burning within our
limitations reached in accuracy and precision of abundance
determinations. In order for our sample stars rotating
that fast, their inclination angles would need to be $i$\,$<$\,10{\degr} for
their typical $v_{\rm rot} \sin i$\,$<$\,30\,km\,s$^{-1}$ (NS11,
NP12), which is statistically unlikely for the majority
of objects.

Observationally, 2/3 of the sample stars are compatible with pristine
N/C ratios (as defined by the cosmic abundance standard, NP12), 
including stars close to the TAMS, implying that many sample stars are
objects with a genuine
$\Omega$/$\Omega_\mathrm{crit}$\,$\lesssim$\,0.4. 
This is in agreement with 
results on a sample of massive detached eclipsing binaries 
\citep[with typical $v_\mathrm{rot}\sin
i$\,$\lesssim$\,100\,km\,s$^{-1}$]{PaHe05,PaSo09,Pavlovskietal09,
Pavlovskietal11,Mayeretal13,Tkachenkoetal14,PaSo14},
which also show no signs of mixing with
CN-cycled matter. Mixing signatures are found in the remaining
1/3 of our sample stars. 
About half of them are compatible with the predictions
from the stellar evolution tracks accounting for rotation
\citep[EGE12;][]{Georgyetal13}. Four objects appear too unevolved to
explain the observed amount of mixing with CN-processed matter. They
may be results of binary evolution \citep{Wellsteinetal01,
deMinketal11}, as discussed below in the notes on individual objects.
The absence of indications for binarity from fitting of their spectral
energy distributions (NP12) imply either a merger or the presence of a
very faint companion.

However, the topic is certainly more complex than apparent at first
glance. For some of the stars it is feasible to determine rotation
periods $P$ from the temporal variability of polarisation
measurements and/or the wind variability established from time-series 
UV spectra \citep[e.g.][]{Moreletal06}.
Coupled with our radius determination (Table~\ref{bigtable}) this 
allows their true $v_\mathrm{rot}$ and the inclination angles $i$ to be
constrained.
Accordingly, 1/5th of the sample stars turn out to be genuine slow rotators,
including several of those showing signatures of CN-mixing compatible 
with initial rotation faster than
$\Omega$/$\Omega_\mathrm{crit}$\,$\approx$\,0.5 (Fig.~\ref{g-NCrot},
stars \#14, 17 and 18). The presence of magnetic fields can apparently 
be considered as an indicator for augmented
mixing in rotating stars \citep[see also][{but see Aerts et al. 2014
for a contrary view.}]{Moreletal08}. Then, angular
momentum losses through a magnetically confined line-driven stellar wind
\citep{Ud-Doulaetal09} seem to be required to produce the observed
slow rotation. The situation in the early B-type stars therefore appears 
to be similar to that of magnetic O stars with observed nitrogen enrichment
\citep{Martinsetal12}. While a qualitative understanding of the
underlying mechanisms may be available, only first steps towards a
comprehensive implementation into stellar evolution models have
been made yet \citep{Meynetetal11}. The probably most fundamental
question that needs clarification is on the origin of the magnetic fields,
whether they are intrinsic or due to some form of dynamo action in the
radiative envelopes of hot, massive stars.

\subsection*{Notes on individual objects}

\paragraph{\#1: HD\,36591.} This star is a member of common proper motion system.
The partner has a $V$ magnitude of 9.8\,mag and is located at a distance
of 2.1{\arcsec} (ET08). It does not contribute any significant light
to the observed spectrum.
\paragraph{\#2: HD\,61068 (PT\,Pup).} The star shows a significantly
enhanced N/C-ratio. Its relatively low age may hint to some form of
binary interaction being responsible for the mixing signature.\\[-8mm]
\paragraph{\#3: HD\,63922 (P\,Pup).} We do not see spectral lines from
the secondary in this binary system in our spectrum, despite it is
relatively bright, $V$\,=\,7.19, and close, at 0.340{\arcsec} distance (ET08).\\[-8mm]
\paragraph{\#4: HD\,74575 ($\alpha$\,Pyx).} \citet{Hubrigetal09}
reported the detection of a magnetic field for this object based 
on spectropolarimetry with {\sc Fors1} on the VLT, which
could not be confirmed by \citet{Bagnuloetal12} by reanalysis of the same 
observational data, and also by \citet{Shultzetal12} based
on spectropolarimetry at high spectral resolution. The star shows a
slightly higher N/C-ratio than the baseline value, which can be 
explained in the framework of rotationally-induced mixing.\\[-8mm]
\paragraph{\#6: HD\,149438 ($\tau$\,Sco).} The star is known to host a
$\sim$500\,G strong magnetic field with a highly complex geometry
\citep{Donatietal06}. Modulation in the spectropolarimetry data and UV stellar wind
line variability point to a period of 41.033$\pm$0.002\,d, which was
identified with the rotational period. With our radius from
Table~\ref{bigtable} a rotational velocity of
$v_\mathrm{rot}$\,=\,6\,km\,s$^{-1}$ is deduced, which implies an
inclination $i$ in the range $\sim$30 to 70{$^{\circ}$} for a
$v_\mathrm{rot} \sin i$\,=\,4$\pm$1\,km\,s$^{-1}$ (NP12). The observed high N/C
ratio could therefore be explained in the scenario of an initially fast
rotator that experienced spin-down from magnetic breaking by mass- and
angular momentum-loss by a line-driven stellar wind \citep{Ud-Doulaetal09}.
This argument apparently fails
in view of the low evolutionary age of $\tau$\,Sco of less than
2\,Myrs (from its position close to the ZAMS, see
Figs.~\ref{teff-logg} or \ref{teff-logL}). However, this evolutionary
age is in conflict with the age of the Upper Scorpius association
\citep[11$\pm$1$\pm$2\,Myr (statistical, systematic),][]{Pecautetal12} 
that $\tau$\,Sco is a member of. $\tau$ Sco is located on the ZAMS
beyond the main sequence turn-off from this association.
The only resort from this is that $\tau$\,Sco is in fact a {\em blue
straggler} that has been rejuvenated either by a merger or by mass
gain in a binary system. The merger scenario is in particular
appealing if the scenario discussed by \citet{Ferrarioetal09}
also works for massive main-sequence mergers \citep[see discussion
by][]{Langer12}, as this would explain the origin of the
observed magnetic field. The current observed slow rotation would then
be the result of angular-momentum loss in a magnetically confined
stellar wind, despite the initial rejection of
the idea (see above).\\[-8mm]
\paragraph{\#7: HD\,886 ($\gamma$\,Peg).} \citet{Handleretal09}
suggest the star to be single from a high-precision space-based
photometric and ground-based spectroscopic observation campaign. This
is in contrast to previous claims, as summarised by ET08, but
consistent with our finding of the absence of second light from the
B2\,IV spectrum. The detection of a magnetic field by 
\citet{BuPl07} appears questionable given the large errors 
of the individual measurements. The non-detection is confirmed by
\citet{Silvesteretal09} and \citet{Neineretal13}.\\[-8mm]
\paragraph{\#8: HD\,29248 ($\nu$\,Eri).}
Asteroseismic analysis of the rich pulsation spectrum of the $\beta$
Cephei star $\nu$\,Eri lead \citet{Pamyatnykhetal04} to conclude that
its equatorial rotation velocity amounts to 
$v_\mathrm{rot}$\,$\approx$\,6\,km\,s$^{-1}$. Their deduced value is 
in contrast to the observed 
$v_\mathrm{rot}\sin i$\,=\,26$\pm$2\,km\,s$^{-1}$ (NP12).
The enhanced N/C-ratio is consistent with predictions from models with 
rotationally induced mixing.\\[-8mm]
\paragraph{\#10: HD\,35708 ($o$\,Tau).} This star is a member of the loose
Cas-Tau association. It is less evolved than the other two Cas-Tau members
of our sample, HD\,3360 and HD\,16582, but also shows an equally high N/C-ratio.
Reaching such a high N/C-ratio early in the evolution would require a
dynamo mechanism to be active and the presence of differential
rotation in the stellar interior, a combination for which a detailed
understanding still has to be developed \citep{Meynetetal11}.
Alternatively, a binary scenario may be invoked to explain the mixing
signature \citep{Langer12, deMinketal13}.\\[-8mm]
\paragraph{\#12: HD\,36822 ($\phi^1$\,Ori).} The star is classified as
SB1, with an orbital period of 3068\,d (ET08). No details are known on
the properties of the secondary, which also does not contribute any
significant second light to our spectrum. The enhanced N/C-ratio is
consistent with predictions from models with rotationally induced mixing.\\[-8mm]  
\paragraph{\#13: HD\,36960.} This is the brightest member of a triple
common proper-motion system. The other two stars, a B1\,V star of $V$-magnitude 
5.65 and a fainter companion at $V$\,=\,8.84, form a tight pair
separated by 36.11{\arcsec} from HD\,36960. This makes the object
effectively a single star for the analysis.\\[-8mm]
\paragraph{\#14: HD\,205021 ($\beta$\,Cep).} \citet{Henrichsetal13}
have recently presented the so far most comprehensive study of the
primary of a triple system. The secondary is a $\Delta V$\,=\,3.4\,mag
fainter Be star, and has an orbital period of $\sim$85\,years.
An angular distance of $\sim$0.25{\arcsec} was found
from speckle interferometry at the epoch of observation \citep{Gezarietal72}. 
We see no indication of second light in the spectrum available to us,
but the Be star is known to produce an emission component in
H$\alpha$ of the primary occasionally, with the spectrum of the
secondary separated using spectro-astrometric techniques \citep{Schnerretal06}.
The third body is an A2\,V star at 7.8\,mag, 13.6{\arcsec} away.
$\beta$\,Cep hosts a sinusoidally varying magnetic field with an amplitude
97$\pm$4\,G and an average value $-$6$\pm$3\,G, implying a polar field
strength of about 300\,G. From the periodicity of the 
UV stellar wind line variability \citet{Henrichsetal13} derived a rotational period of 
$P$\,=\,12.00075(11)\,d (which is compatible with the magnetic
field signal). This implies a true rotational velocity of
$v_\mathrm{rot}$\,=\,24$\pm$2\,km\,s$^{-1}$, using our radius value 
(see Table~\ref{bigtable}). This is compatible with the spectroscopically 
determined value $v_\mathrm{rot}\,\sin i$\,=\,28$\pm$3\,km\,s$^{-1}$
(NP12), requiring that the star is seen close to equator-on.
Despite this slow rotation, the star shows high N/C. It may be speculated 
that initially this was a faster rotator, thus
explaining the mixing of the surface with CN-processed material, while 
magnetic breaking due to angular momentum losses by a magnetically confined 
line-driven stellar wind has lead to the spin-down \citep{Ud-Doulaetal09}.
Overall, there is excellent agreement of $\beta$\,Cep's fundamental
parameters derived by \citet{Henrichsetal13} and our values, 
despite somewhat different atmospheric parameters. Note that the
$M_V^0$\,=\,$-$5.8$\pm$0.2 given by \citet{Henrichsetal13} is probably
a typo.\\[-8mm]
\paragraph{\#16: HD\,216916 (EN Lac).} The star
belongs to a triple system. The B2\,IV primary forms a close pair with
the secondary of spectral type F6-7 with a 12.10\,d period. The third
body in the system is an 11.4\,mag F0 star at 27.6{\arcsec} distance (ET08).
The two other objects do not contribute light to the spectrum of the
primary in a significant way, making it effectively a single star in
terms of interpretation.\\[-8mm]
\paragraph{\#17: HD\,3360 ($\zeta$\,Cas).} \citet{Neineretal03} find
this member of the Cas-Tau association
to be magnetic on the basis of spectropolarimetric
observations, with an inferred polar field strength of 335\,G. From the
periodicity of the magnetic signal and UV stellar wind line variability 
they derived a rotational period of 
$P$\,=\,5.37\,d, which results in a true rotational velocity of 
$v_\mathrm{rot}$\,=\,59$\pm$5\,km\,s$^{-1}$, employing our radius value
(see Table~\ref{bigtable}). From an observed projected rotational
velocity of $v_\mathrm{rot}\sin i$\,=\,20$\pm$2\,km\,s$^{-1}$ (NP12) follows
$i$\,$\approx$\,20{\degr}. The star shows one of the highest N/C
ratios in our sample, despite being a slow rotator.
It may be speculated that initially this was a faster rotator, thus
explaining the mixing of the surface with CN-processed material, while 
magnetic breaking due to angular momentum losses by a magnetically confined 
line-driven stellar wind has lead to the current low
rotation rate over its $\sim$25\,Myr lifetime \citep[which is within a
factor $\sim$2.5 of the prediction by][]{Ud-Doulaetal09}. With such an age
$\zeta$\,Cas is close to the TAMS. There is excellent
agreement between all stellar parameters presented here and those
derived by \citet{Neineretal03}.\\[-8mm]
\paragraph{\#18: HD\,16582 ($\delta$\,Cet).} The star resembles
closely HD\,3360 in almost all aspects (see Table~\ref{bigtable}),
including its membership to the Cas-Tau association.
It is a true slow rotator with $v_\mathrm{rot}$\,=\,14 or
28\,km\,s$^{-1}$, as suggested by asteroseismic analysis
\citep{Aertsetal06}. From the observed $v_\mathrm{rot}\sin
i$\,=\,15$\pm$2\,km\,s$^{-1}$ (NP12) we can reject the second solution
of \citet{Aertsetal06} and conclude that the star is seen equator-on,
with a rotational period of about 22\,d (employing $R$ from Table~\ref{bigtable}).
\citet{Hubrigetal09} reported the detection of a magnetic field for this object 
based on spectropolarimetric observations with {\sc Fors1} on the VLT, which
could not be confirmed by \citet{Bagnuloetal12} by reanalysis
of the same observational data, and also by \citet{Silvesteretal09} based
on spectropolarimetry at high spectral resolution. 
The origin of the mixing of the surface with CN-cycled material
is compatible with evolution models accounting for the effects of rotation.
However, an initial $\Omega/\Omega_\mathrm{crit}$ of
the order 0.7 is required (see Fig.~\ref{g-NCrot}), and therefore leaves open the
question how the star has lost practically its entire angular
momentum. One may speculate about a magnetic field generated by a
dynamo and magnetic braking, which suppressed the dynamo successively.
Alternatively, a binary scenario may be invoked to explain the mixing
signature \citep{Langer12, deMinketal13}.\\[-8mm] 
\paragraph{\#20: HD\,160762 ($\iota$\,Her).} The B3\,IV star is the
primary in a binary, with a companion of unknown spectral type. The
orbital period is 113.8\,d (ET08). We see no indication of
lines produced by the secondary in the spectrum available to us.\\[-8mm]
\paragraph{\#21: HD\,37020 ($\theta^1$\,Ori\,A).}
This star of the Orion Trapezium is an eclipsing binary 
\citep[for a discussion see][]{StLl00}, consisting of the B0.5\,V
primary and a low-luminosity companion (contributing no significant
second light to our spectrum), a probably late-type pre-main 
sequence star. The eclipses imply $i$\,$\approx$\,90{\degr}, therefore 
$v_\mathrm{rot}$\,=\,45$\pm$3\,km\,s$^{-1}$, adopting $v \sin i$ from
NS11. No indication of mixing with CN-cycled products is found for
this slow rotator on the ZAMS.\\[-8mm]
\paragraph{\#22: HD\,37042 ($\theta^2$\,Ori\,B).} A rv-variable
\citep{MoLe91}. We find no indication of second light in our
spectrum. Only a binary merger seems to be able to produce such a high N/C-ratio 
that near the ZAMS (see the discussion on $\tau$\,Sco above).
The merger should be a fast-spinning object initially, requiring
HD\,37042 to be seen nearly pole-on to produce the observed
$v_\mathrm{rot} \sin i$\,=\,30\,km\,s$^{-1}$ (NP12).\\[-8mm]
\paragraph{\#23: HD\,36959.} A potential rv-variable
\citep{MoLe91}. We find no indication of second light in our spectrum.\\[-8mm]
\paragraph{\#26: HD\,35039 ($o$\,Ori).} Periodic rv-variations and the
non-detection of spectral signatures from a second object lead to a
classification as a single-lined spectroscopic binary (SB1). The
secondary is of unknown spectral type, and the system has a period of
290\,days (ET08). Based on its location in the HRD, the star is near
the TAMS and may even have terminated core-hydrogen burning already.


\begin{figure}[!t]
\centering
\includegraphics[width=.99\linewidth]{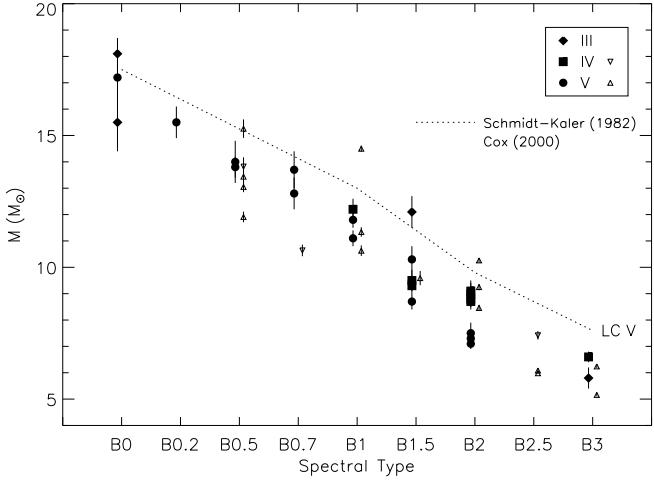}
\caption{Masses of the sample stars (full symbols) as a function of 
spectral type. Encoding of the luminosity class according to the
legend, open triangles denote data derived from detached eclipsing 
binary stars \citep{Torresetal10}. Error bars denote
$1\sigma$-uncertainties. The functional relationship for dwarfs
(luminosity class V) as advocated in the astrophysical reference 
literature -- Landolt-B\"ornstein \citep{Schmidt-Kaler82}, Allen's
Astrophysical Quantities \citep{Cox00} -- is also indicated.  
}
\label{SpT-M}
\end{figure}

\begin{figure}[!t]
\centering
\includegraphics[width=.99\linewidth]{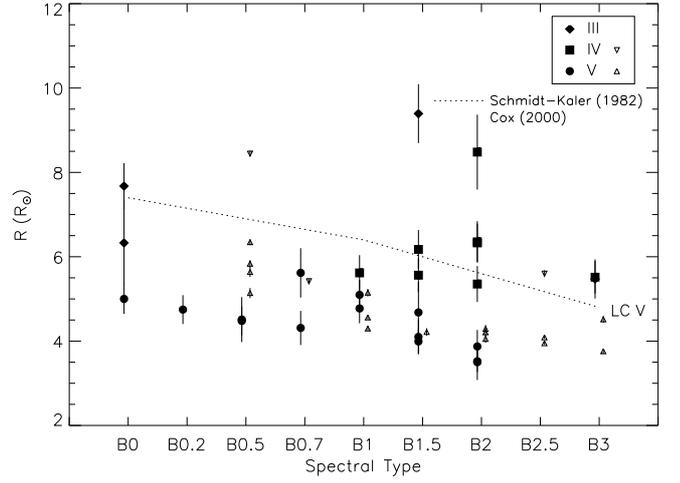}
\caption{Radii of the sample stars as a function of spectral type,
see Fig.~\ref{SpT-M} for further explanation.
}
\label{SpT-R}
\end{figure}

\section{Functional relationships\label{sectfunctional}}
The good agreement of our and the binary data in terms of the $M$-$R$
and $M$-$L$ relationships motivates us to perform further
comparisons. The aim of this section is to verify data from the
astrophysical reference literature, from the Landolt-B\"ornstein \citep{Schmidt-Kaler82} 
which also provides the basis for the values presented in
Allen's Astrophysical Quantities \citep{Cox00}. 

Masses of the sample stars as a function of spectral type and luminosity 
class are discussed in Fig.~\ref{SpT-M}. As expected, there is an overall 
good match between our and the DEB data. Within a spectral class, giants appear 
to have higher masses than subgiants by tendency, which in turn appear to have
higher masses than dwarfs, though the number statistics for the more
evolved objects is arguably small. On the other hand, the reference
literature relation is clearly off, indicating masses systematically
higher by $\sim$10-20\%. 

\begin{figure}[!t]
\centering
\includegraphics[width=.99\linewidth]{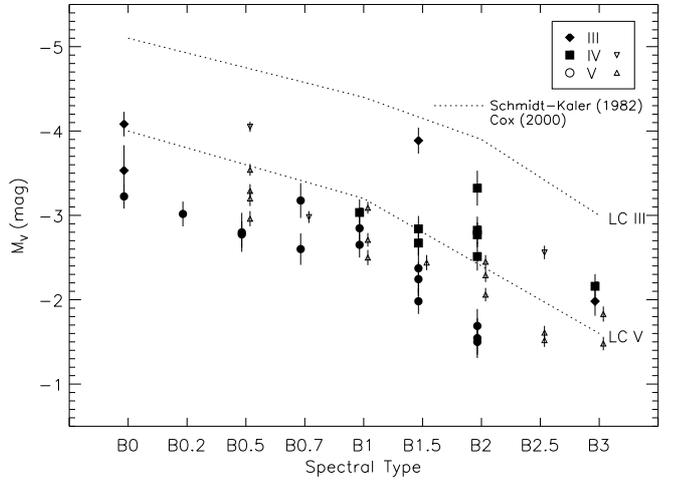}
\caption{Absolute visual magnitudes of the sample stars as a function
of spectral type, see Fig.~\ref{SpT-M} for further explanation. 
The functional relationship for giants (luminosity class III) from the
reference literature is also indicated.
}
\label{SpT-MV0}
\end{figure}

\begin{figure}[!t]
\centering
\includegraphics[width=.99\linewidth]{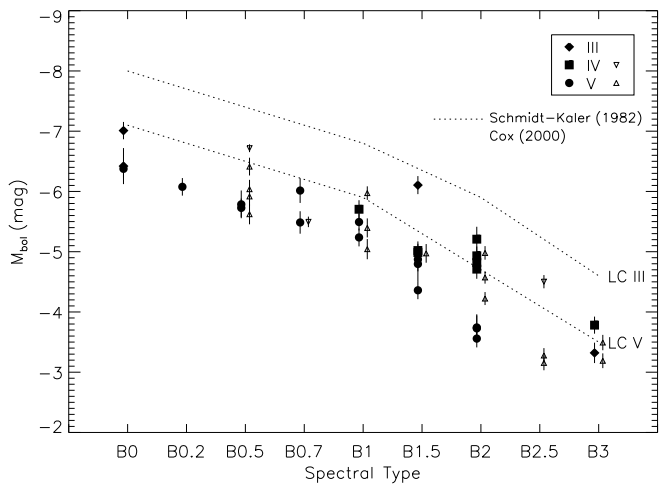}
\caption{Absolute bolometric magnitudes of the sample stars as a
function of spectral type, see Fig.~\ref{SpT-M} for further explanation.
The functional relationship for giants (luminosity class III) from the
reference literature is also indicated.
}
\label{SpT-Mbol}
\end{figure}

A similar picture is found for radii as a function of spectral type
(Fig.~\ref{SpT-R}): our and the DEB data agree, and fall systematically 
below the reference relation by $\sim$25\% 
for luminosity
class V. As expected, radii become larger from dwarfs to giants.
Note that in terms of observational quantities, $g \propto M/R^2$.
Consequently, our data implies an average $\log g$\,$\approx$\,4.15-4.20 for 
luminosity class V, while the \citet{Schmidt-Kaler82} relation
indicates $\log g$\,$\approx$3.95, more appropriate for luminosity
class IV in our analysis. 

A direct consequence of our smaller masses and radii and therefore
higher gravities is that the stars are more compact and significantly fainter than
assumed in the reference literature. The difference in absolute visual
and bolometric magnitude is $\sim$0.5\,mag, see Figs.~\ref{SpT-MV0}
and \ref{SpT-Mbol}, respectively. Again, this is in agreement with the DEB data.
Note that a discussion of the $T_\mathrm{eff}$--spectral-type relation is omitted 
here as this has been investigated in detail by \citet{Nieva13}.

The reasons for the discrepancies between the relations implied by our
data and the modern DEB data on one hand, and the data from the reference
literature on the other hand appear to have a historical background. 
Regarding masses and radii, it is clear from the critical evaluation of
the most reliable data from binary stars by \citet[including his own
huge efforts]{Popper80}, and subsequent critical compilations 
by \citet{HaHe81} and \citet{UnDo82}, how uncertain and scarce reliable data 
for OB stars were at the time of the compilation of the
Landolt-B\"ornstein input \citep{Schmidt-Kaler82}. The situation was improving 
only by observations made with the first electronic detectors. This is
clear from the compilations of \citet{Harmanec88} and later of
\citet{Andersen91}. These tabulations of masses and radii 
are already in rather good agreement with the present data.

In the case of absolute visual and bolometric magnitudes additional reasons 
stem from systematics in the distance scale to star clusters and associations,
and to differences in bolometric correction scales. The
high frequency of multiples among massive OB-stars has been
uncovered only recently, reaching 50--70\% \citep{Chinietal12,Sanaetal12}.
As a consequence, higher luminosities for OB stars may have resulted
on average in the past because of a contribution from binaries mistaken 
for single stars. 

We conclude that the general astrophysics reference literature on
OB-type stars needs to be updated with regard to masses, radii and luminosities 
as a function of spectral type. As the differences of modern results and established
data from the reference literature will extend beyond the early B-type
stars investigated here, a more comprehensive study than the present
one is clearly required. However, we refrain here from providing
reference values from our sample as a first step in that direction
because many spectral type/luminosity class combinations are
covered only by one or no observed data point, in particular for
subgiants and giants. It is therefore not feasible to give average
values or ranges in a meaningful way. We postpone this to the near
future, when larger samples of accurate and precise parameters of OB stars
will become available. For the moment Table~\ref{bigtable}
allows all relevant data in this context to be extracted.


\section{Concluding remarks}\label{conclusion}
We have shown that on the basis of a careful
non-LTE analysis of high-quality spectra, coupled to state-of-the-art
stellar evolution models, accurate and precise
fundamental parameters of single core hydrogen-burning early B-type 
stars can be determined.
Evolutionary masses, radii, and luminosities can thus be constrained to better 
than typically 5\%, 10\%, and 20\% uncertainty, respectively, coming
to within a factor 2--3 of data derived from the best
indicators, detached eclipsing binaries \citep{Torresetal10}. 
Spectroscopic masses show uncertainties by a factor 5 to 10
higher, which is only because of the highly uncertain distances. There is good
agreement of evolutionary tracks from different modelling approaches 
for slowly-/non-rotating massive stars on the main sequence
\citep{MaPa13}, such that the low uncertainties of evolutionary masses
appear realistic.

Good quantitative agreement
of the single star and DEB properties is found, while differences to
accepted values from the reference literature \citep{Schmidt-Kaler82,
Cox00} are noted. An extension of the present work is clearly required to 
update the reference values over a larger spectral range. 
Objects that are influenced by second light in the photometric and
spectral data need carefully to be rejected (except for DEBs). The determination of the
fundamental parameters of the bona-fide single stars can then benefit from 
the progress made in stellar atmosphere modelling over the past three
decades, as shown here.

The vast majority of our sample early B-type stars from the solar
neighbourhood appear to be
genuine slow rotators. Their observational properties are overall in 
agreement with the recent Geneva stellar evolution models by
\citet{Ekstroemetal12} and
\citet{Georgyetal13}. A few stars are found that seem incompatible
with predicted characteristics. One group is that of somewhat evolved genuine slow
rotators that show signatures of mixing of CN-processed matter. These
may be explained by magnetically augmented mixing in an initially
faster rotator that was spun-down by angular momentum losses through a
magnetically confined stellar wind \citep{Meynetetal11}. Another group are apparently young
but already significantly nitrogen-enriched stars that may stem from a
binary interaction scenario \citep{deMinketal11}. In particular, $\tau$\,Sco turns out to
be a blue straggler, rejuvenated possibly via a merger, which may also
explain the origin of its magnetic field according to the mechanism
suggested by \citet{Ferrarioetal09}.


\begin{acknowledgements}
The authors would like to thank the staff of OAC \& IATE at
Universidad Nacional de C\'ordoba for their hospitality during their 
stay where part of the present work was accomplished.
We also thank S.~Kimeswenger for stimulating discussion and the
anonymous referee for valuable comments.
MFN acknowledges financial support by the equal opportunities program
FFL of the University of Erlangen-Nuremberg.
\end{acknowledgements}



\begin{thebibliography}{}

\bibitem[Aerts et al.(2014)]{Aertsetal14}
Aerts, C., Molenberghs, G., Kenward, M. G., \& Neiner, C. 2014, \apj, 781, 88

\bibitem[Aerts et al.(2006)]{Aertsetal06}
Aerts, C., Marchenko, S. V., Matthews, J. M., et al. 2006, \apj, 642, 470

\bibitem[Andersen(1991)]{Andersen91}
Andersen, J. 1991, \aapr, 3, 91

\bibitem[Asplund et al.(2009)]{AGSS09}
Asplund, M., Grevesse, N., Sauval, A.~J., \& Scott, P.~2009, \araa, 47, 481

\bibitem[Bagnulo et al.(2006)]{Bagnuloetal06}
Bagnulo, S., Landstreet, J. D., Mason, E., et al. 2006, \aap, 450, 777

\bibitem[Bagnulo et al.(2012)]{Bagnuloetal12}
Bagnulo, S., Landstreet, J. D., Fossati, L., \& Kochukhov, O. 2012, \aap, 538, A129 

\bibitem[Bally(2008)]{Bally08}
Bally, J. 2008, in Handbook of Star Forming Regions, Volume I: The
Northern Sky. ed. B. Reipurth (San Francisco: ASP), 459

\bibitem[Blaauw(1958)]{Blaauw58}
Blaauw, A. 1958, \aj, 63, 186

\bibitem[Blaauw(1991)]{Blaauw91}
Blaauw, A. 1991, in The Physics of Star Formation and Early Stellar
Evolution, ed. C. J. Lada \& N. D. Kylafis (Dordrecht: Kluwer), 125

\bibitem[Bresolin et~al.(2008)]{Bresolinetal08}
Bresolin, F., Crowther, P.~A., \& Puls, J., eds. 2008, IAU Symposium,
Vol. 250, {Massive Stars as Cosmic Engines}

\bibitem[Brott et al.(2011)]{Brottetal11}
Brott, I., de Mink, S. E., Cantiello, M., et al. 2011, \aap, 530, A115

\bibitem[Butkovskaya \& Plachinda(2007)]{BuPl07}
Butkovskaya, V. V., \& Plachinda, S. I. 2007, \aap, 469, 1069

\bibitem[Butler \& Giddings(1985)]{BuGi85}
Butler, K., \& Giddings, J. R.~1985, in
Newsletter of Analysis of Astronomical Spectra, No. 9 (Univ. London)

\bibitem[Chieffi \& Limongi(2013)]{ChLi13}
Chieffi, A., \& Limongi, M. 2013, \apj, 764, 21

\bibitem[Chini et al.(2012)]{Chinietal12}
Chini, R., Hoffmeister, V.~H., Nasseri, A., Stahl, O., \& Zinnecker, H. 2012, \mnras, 424, 1925

\bibitem[Cox(2000)]{Cox00}
Cox, A. N. 2000, Allen’s astrophysical quantities, ed. Cox, A. N. (Springer-Verlag, New York)

\bibitem[de Mink et al.(2011)]{deMinketal11}
de Mink, S. E., Langer, N., \& Izzard, R. G. 2011, Bull.              
Soc. R. Sci. Liege, 80, 543

\bibitem[de Mink et al.(2013)]{deMinketal13}
de Mink, S. E., Langer, N., Izzard, R. G., et al. 2013, \apj, 764, 166

\bibitem[Donati et al.(2006)]{Donatietal06}
Donati, J.-F., Howarth, I. D., Jardine, M. M., et al. 2006, \mnras, 370, 629

\bibitem[Eggleton \& Tokovinin(2008)]{EgTo08}
Eggleton, P. P., \& Tokovinin, A. A. 2008, \mnras, 389, 869 (ET08)

\bibitem[Ekstr\"om et al.(2008)]{Ekstroemetal08}
Ekstr\"om, S., Meynet, G., Maeder, A., \& Barblan, F. 2008, \aap, 478, 467

\bibitem[Ekstr\"om et al.(2012)]{Ekstroemetal12}
Ekstr\"om, S., Georgy, C., Eggenberger, P., et al. 2012, \aap, 537, A146 (EGE12)

\bibitem[Esteban et al.(2004)]{Estebanetal04}
Esteban, C., Peimbert, M., Garc\'ia-Rojas, J., et al. 2004, \mnras, 355, 229

\bibitem[Ferrario et al.(2009)]{Ferrarioetal09}
Ferrario, L., Pringle, J. E., Tout, C. A., \& Wickramasinghe, D. T. 2009,
\mnras, 400, L71

\bibitem[Georgy et al.(2013)]{Georgyetal13}
Georgy, C., Ekstr\"om, S., Granada, A., et al. 2013, \aap, 553, A24

\bibitem[Gezari et al.(1972)]{Gezarietal72}
Gezari, D. Y., Labeyrie, A., \& Stachnik, R. V. 1972, \apj, 173, L1

\bibitem[Giddings(1981)]{Giddings81}
Giddings, J. R.~1981, Ph.D. Thesis, University of London

\bibitem[Gies \& Lambert(1992)]{GiLa92}
Gies, D. R., \& Lambert, D. L. 1992, \apj, 387, 673

\bibitem[Gulliver et al.(1994)]{Gulliveretal94}
Gulliver, A. F., Hill, G., \& Adelman, S. J. 1994, \apj, 429, L81

\bibitem[Habets \& Heintze(1981)]{HaHe81}
Habets, G. M. H. J., \& Heintze, J. R. W. 1981, \aaps, 46, 193

\bibitem[Handler et al.(2009)]{Handleretal09}
Handler, G., Matthews, J. M., Eaton, J. A., et al. 2009, 698, L56

\bibitem[Harmanec(1988)]{Harmanec88}
Harmanec, P. 1988, Bull. Astron. Inst. Czechoslovakia, 39, 329

\bibitem[Heger \& Langer(2000)]{HeLa00}
Heger, A., \& Langer, N. 2000, \apj, 544, 1016

\bibitem[Heger et al.(2005)]{Hegeretal05}
Heger, A., Woosley, S. E., \& Spruit, H. C. 2005, \apj, 626, 350

\bibitem[Henrichs et al.(2013)]{Henrichsetal13}
Henrichs, H. F., de Jong, J. A., Verdugo, E., et al. 2013, \aap, 555, A46

\bibitem[Hirschi et al.(2004)]{Hirschietal04}
Hirschi, R., Meynet, G., \& Maeder, A. 2004, \aap, 425, 649

\bibitem[Hohle et al.(2010)]{Hohleetal10}
Hohle, M. M., Neuh\"auser, R., \& Schutz, B. F. 2010, Astron.~Nachr., 331, 349

\bibitem[Hubrig et al.(2009)]{Hubrigetal09}
Hubrig, S., Briquet, M., De Cat, P., et al. 2009, Astron. Nachr., 330, 317

\bibitem[Kilian(1992)]{Kilian92}
Kilian, J. 1992, \aap, 262, 171

\bibitem[Kurucz(1993)]{Kurucz93}
Kurucz, R. L.~1993, CD-ROM No. 13 (SAO, Cambridge, Mass.)

\bibitem[Langer(2012)]{Langer12}
Langer, N.  2012, \araa, 50, 107

\bibitem[Lyubimkov et al.(2002)]{Lyubimkovetal02}
Lyubimkov, L. S., Rachkovskaya, T. M., Rostopchin, S. I., \& Lambert, D. L. 2002, \mnras, 333, 9

\bibitem[Maeder \& Conti(1994)]{MaCo94}
Maeder, A., \& Conti, P.~S. 1994, \araa, 32, 227

\bibitem[Maeder \& Meynet(2005)]{MaMe05}
Maeder, A., \& Meynet, G. 2005, \aap, 440, 1041

\bibitem[Maeder \& Meynet(2012)]{MaMe12}
Maeder, A., \& Meynet, G. 2012, Rev. Mod. Phys., 84, 25

\bibitem[Maeder et al.(2014)]{Maederetal14}
Maeder, A., Przybilla, N., Nieva, M. F., et al. 2014, \aap, 565, A39

\bibitem[Martins \& Palacios(2013)]{MaPa13}
Martins, F., \& Palacios, A. 2013, \aap, 560, A16

\bibitem[Martins et al.(2012)]{Martinsetal12}
Martins, F., Escolano, C., Wade, G. A., et al. 2012, \aap, 538, A29 

\bibitem[Mayer et al.(2013)]{Mayeretal13}
Mayer, P., Harmanec, P., \& Pavlovski, K. 2013, \aap, 550, A2

\bibitem[Menten et al.(2007)]{Mentenetal07}
Menten, K. M., Reid, M. J., Forbrich, J., \& Brunthaler, A. 2007, \aap, 474, 515

\bibitem[Meynet \& Maeder(2000)]{MeMa00}
Meynet, G., \& Maeder, A. 2000, \aap, 361, 101

\bibitem[Meynet \& Maeder(2003)]{MeMa03}
Meynet, G. \& Maeder, A.~2003, \aap, 411, 543

\bibitem[Meynet \& Maeder(2005)]{MeMa05}
Meynet, G. \& Maeder, A. 2005, \aap, 429, 581

\bibitem[Meynet et al.(2011)]{Meynetetal11}
Meynet, G., Eggenberger, P., \& Maeder, A. 2011, \aap, 525, L11

\bibitem[Morel et al.(2006)]{Moreletal06}
Morel, T., Butler, K., Aerts, C., et al. 2006, \aap, 457, 651

\bibitem[Morel et al.(2008)]{Moreletal08}
Morel, T., Hubrig, S., \& Briquet, M. 2008, \aap, 481, 453

\bibitem[Morrell \& Levato(1991)]{MoLe91}
Morrell, N., \& Levato, H. 1991, \apjs, 75, 965 

\bibitem[Moultaka et al.(2004)]{Moultakaetal04}
Moultaka, J., Ilovaisky, S. A., Prugniel, P., \& Soubiran, C. 2004, \pasp, 116, 693

\bibitem[Neiner et al.(2003)]{Neineretal03}
Neiner, C., Geers, V. C., Henrichs, H. F., et al. 2003, \aap, 406, 1019

\bibitem[Neiner et al.(2014)]{Neineretal13}
Neiner, C., Monin, D., Leroy, B., Mathis, S., \& Bohlender, D. 2014, \aap, 562, A59

\bibitem[Nieva(2013)]{Nieva13}
Nieva, M.~F. 2013, \aap, 550, A26

\bibitem[Nieva \& Przybilla(2006)]{NP06}
Nieva, M.~F., \& Przybilla,~N.~2006, \apj, 639, L39

\bibitem[Nieva \& Przybilla(2007)]{NP07}
Nieva, M. F., \& Przybilla,~N.~2007, \aap, 467, 295

\bibitem[Nieva \& Przybilla(2008)]{NP08}
Nieva, M. F., \& Przybilla, N.~2008, \aap, 481, 199

\bibitem[Nieva \& Przybilla(2012)]{NP12}
Nieva, M. F., \& Przybilla, N. 2012, \aap, 539, A143 (NP12)

\bibitem[Nieva \& Sim\'on-D\'iaz(2011)]{NS11}
Nieva, M. F., \& Sim\'on-D\'iaz, S. 2011, \aap, 532, A2 (NS11)

\bibitem[Pamyatnykh et al.(2004)]{Pamyatnykhetal04}
Pamyatnykh, A. A., Handler, G., \& Dziembowski, W. A. 2004, \mnras, 350, 1022

\bibitem[Pavlovski \& Hensberge(2005)]{PaHe05}
Pavlovski, K., \& Hensberge, H. 2005, \aap, 439, 309

\bibitem[Pavlovski \& Southworth(2009)]{PaSo09}
Pavlovski, K., \& Southworth, J. 2009, \mnras, 394, 1519

\bibitem[Pavlovski \& Southworth(2014)]{PaSo14}
Pavlovski, K., \& Southworth, J. 2014, EAS Publ.~Ser., 64, 29

\bibitem[Pavlovski et al.(2011)]{Pavlovskietal11}
Pavlovski, K., Southworth, J., Tamajo, E., \& Kolbas, V. 2011, Bull.
Soc. R. Sci. Liege, 80, 714

\bibitem[Pavlovski et al.(2009)]{Pavlovskietal09}
Pavlovski, K., Tamajo, E., Koubsk\'y, P., et al. 2009, \mnras, 400, 791

\bibitem[Pecaut et al.(2012)]{Pecautetal12}
Pecaut, M. J., Mamajek, E. E., \& Bubar, E. J. 2012, \apj, 746, 154

\bibitem[Popper(1980)]{Popper80}
Popper, D. M.1980, \araa, 18, 115

\bibitem[Przybilla et al.(2008)]{Przybillaetal08}
Przybilla, N., Nieva, M. F., \& Butler, K. 2008, \apj, 688, L103

\bibitem[Przybilla et al.(2010)]{Przybillaetal10}
Przybilla, N., Firnstein, M., Nieva, M. F., Meynet, G., \& Maeder, A. 2010, \aap, 517, A38

\bibitem[Przybilla et al.(2011)]{PNB11}
Przybilla, N., Nieva, M. F., \& Butler, K.~2011, J.~Phys.: Conf.~Ser.,
328, 012015

\bibitem[Sana et al.(2012)]{Sanaetal12}
Sana, H., de Mink, S.~E., de Koter, A., et~al. 2012, Science, 337, 444

\bibitem[Schmidt-Kaler(1982)]{Schmidt-Kaler82}
Schmidt-Kaler, T. 1982, in: Landolt-B\"ornstein, Vol. 2, Subvol. b (Springer-Verlag, Berlin)

\bibitem[Schnerr et al.(2006)]{Schnerretal06}
Schnerr, R. S., Henrichs, H. F., Oudmaijer, R. D., \& Telting, J. H.
2006, \aap, 459, L21

\bibitem[Schnerr et al.(2008)]{Schnerretal08}
Schnerr, R. S., Henrichs, H. F., Neiner, C., et al. 2008, \aap, 483, 857

\bibitem[Shultz et al.(2012)]{Shultzetal12}
Shultz, M., Wade, G. A., Grunhut, J., et al. 2012, \apj, 750, 2

\bibitem[Silvester et al.(2009)]{Silvesteretal09}
Silvester, J., Neiner, C., Henrichs, H. F., et al. 2009, \mnras, 398, 1505

\bibitem[Song et al.(2013)]{Songetal13}
Song, H. F., Maeder, A., Meynet, G., et al. 2013., \aap, 556, A100

\bibitem[Strickland \& Lloyd(2000)]{StLl00}
Strickland, D. J., \& Lloyd, C. 2000, Observatory, 120, 141

\bibitem[Tkachenko et al.(2014)]{Tkachenkoetal14}
Tkachenko, A., Degroote, P., Aerts, C., et al. 2014, \mnras, 438, 3093

\bibitem[Torres et al.(2010)]{Torresetal10}
Torres, G., Andersen, J., \& Gim\'enez, A. 2010, \aapr, 18, 67 (TAG10)

\bibitem[ud-Doula et al.(2009)]{Ud-Doulaetal09}
Ud-Doula, A., Owocki, S. P., \& Townsend, R. H. D. 2009, \mnras, 392, 1022

\bibitem[Underhill \& Doazan(1982)]{UnDo82}
Underhill, A., \& Doazan, V. 1982, B Stars with and without emission
lines, NASA-SP--456 (NASA, Washington, D.C.)

\bibitem[van Leeuwen(2007)]{vanLeeuwen07}
van Leeuwen, F. 2007, Hipparcos, the New Reduction of the Raw Data
(Springer-Verlag, Berlin)

\bibitem[van Leeuwen(2009)]{vanLeeuwen09}
van Leeuwen, F. 2009, \aap, 497, 209

\bibitem[Vink et al.(2001)]{Vinketal01}
Vink, J. S., de Koter, A., \& Lamers, H. J. G. L. M. 2001, \aap, 369, 574

\bibitem[Vanbeveren et~al.(1998)]{Vanbeverenetal98}
Vanbeveren, D., De Loore, C., \& Van Rensbergen, W. 1998, \aapr, 9, 63

\bibitem[Wellstein et al.(2001)]{Wellsteinetal01}
Wellstein, S., Langer, N., \& Braun, H. 2001, \aap, 369, 939

\bibitem[Wolff(1990)]{Wolff90}
Wolff, S. C. 1990, \aj, 100, 1994

\bibitem[Woosley \& Heger(2012)]{WoHe12}
Woosley, S.~E., \& Heger, A. 2012, \apj, 752, 32

\bibitem[de Zeeuw et al.(1999)]{deZeeuwetal99}
de Zeeuw, P. T., Hoogerwerf, R., de Bruijne, J. H. J., Brown, A. G. A., \& Blaauw, A. 
1999, \aj, 117, 354

\end{thebibliography}
\end{document}